\documentclass{article}
\usepackage{graphicx}
\def\s.) 1.) 2 {\langle.)1,.) 2\rangle}
\def\bb{\begin{equation}}
\def\ee{\end{equation}}
\def\th.)1{\noindent{\bf.)1}
\hspace{1mm}}

\def\Re{\hbox{Re}}
\def\ve{\varepsilon}
\def\s{\sigma}
\def\a{\alpha}
\def\b{\beta}
\def\d{\delta}
\def\k{\kappa}
\def\pt{\partial}
\def\matrix{\begin{array}}
\def\endmatrix{\end{array}}

\title{
The solution of the Painleve equations as special functions of
catastrophes, defined by a rejection in these equations of terms with
derivative}
\author{O.M. Kiselev,  B.I.  Suleimanov\\
Institute of Mathematics,  Ufa Sci. Centre, of  Russian Acad. of
Sci.\\ Institute of Mathematics, 112,Chernyshevsky str., Ufa,
\\450000, Russia\\
E-mail: ok@imat.rb.ru  bis@imat.rb.ru}
\date{February, 4, 1999}
\begin{document}
\maketitle

%------------------------- A B S T R A C T --------------------------
\begin{abstract}
The relation between the Painleve equations and the algebraic
equations with the catastrophe theory point of view are considered.
The asymptotic solutions with respect to the small parameter of the
Painleve equations different types are discussed. The qualitative
analysis of the relation between algebraic and fast oscillating
solutions is done for Painleve-2 as an example.
\end{abstract}

\vspace{5mm}
\par
1. From six equations the Painleve ($ a $, $ b $, $ c $, $ d $ - constants)

\begin{eqnarray*}
P1:&&w_{xx}=6w^2+x,\\ P2:&&w_{xx}=2w^3+xv+a,\\ P3:&&w_{xx}={w^2_x\over
w}-{w_x\over x}+{(aw^2+b)\over x}+cw^3+{d\over w},\\
P4:&&w_{xx}={w^2_x\over {2w}}+{3w^3\over 2}+4xw^2+2(x^2-a)w+{8b\over
w},\\ P5:&&w_{xx}=({1\over {2w}} +{1\over {w-1}})w^2_x-{w_x\over
x}+{2\over x^2}(w-1)^2({aw+b\over w})+{cw\over x}+{dw(w+1)\over
{w-1}},\\ P6:&&w_{xx}={1\over 2}({1\over {w}}+{1\over {w-1}}+ {1\over
{w-x}} )w^2_x
-({1\over x}+{1\over {x-1}}+1\ {w-x})w_x+
\\
&&{w(w-1)(w-x)\over {x^2(x-1)^2}}[{a+bx\over w^2}+
{c\over (w-1)^2}(x-1)+{dx(x-1)\over (w-x)^2}],
\end{eqnarray*}

first five equations can be obtained from sixth using  following passages
to the limit \cite{A}:

substitution in $ P6 $ $ 1 + {\epsilon} x $ instead of $ x $, $ d /
{\epsilon}^2 $ instead of $ d $, $ c {\epsilon} $ - $ d {\epsilon}^2 $
instead of $ c $ with $ \epsilon $ tending to zero gives the equation $ P5
$;

in the equation $ P5 $ a substitution $ 1 + {\epsilon} w $ instead of $ w
$, $ -b / {\epsilon} ^ 2 $ instead of $ b $, $ b / {\epsilon} ^ 2 $ + $ a /
{\epsilon} $ instead of $ a $, $ c / {\epsilon} $ instead of $ c $ $ d /
{\epsilon} $ instead of $ d $ and the passage to the limit $ {\epsilon} \to
{0} $ gives the equation $ P3 $;

in the same fifth equation the Painleve we shall substitute $ {\epsilon} w
{\sqrt{2}} $ instead of $ w $, $ 1 + {\epsilon} {\sqrt{2}} $ instead of $ w
$, $ 1 / (2 {\epsilon}^4) $ instead of $ a $, $ -1 / {\epsilon}^4 $ instead
of $ c $, $ -1 / (2 {\epsilon}^4) $ $ d / {\epsilon} $ instead of $ d $. In
this case in a limit there will be an equation $ P3 $;

in turn, substitution in equation $ P3 $ $ 1 + {\epsilon}^2x $ instead
of $ x $, $ 1 + 2 {\epsilon} w $ instead of $ w $, $ 1 / (4
{\epsilon}^6) $ instead of $ c $, - $ 1 / (2 {\epsilon}^6) $ instead
of $ a $, $ 1 / (2 {\epsilon}^6 $ + $ 2b / {\epsilon}^3 $ instead of $
b $ gives in a limit the equation $ P2 $;

the equation $ P2 $ can be obtained by a passage to the limit as well from
the equation $ P4 $ using a substitution $ {\epsilon} x2^{-1/3} $ - $ 1
/ {\epsilon}^3 $ instead of $ x $, $ 2^{2/3} {\epsilon} w $ + $ 1 /
{\epsilon}^3 $ instead of $ w $, $ -1 / (2 {\epsilon}^6) -a $ instead of $
a $, $ -1 / (2 {\epsilon}^{12}) $ instead of $ b $;

and, at last, $ P1 $ it turns out as a limit from the equation $ P2 $ at a
substitution ${\epsilon}^2x - 6 / {\epsilon}^{10} $ instead of $x$, $
{\epsilon} w + 1 / {\epsilon}^5 $ instead of $ w $, $ 4 / {\epsilon}^{15} $
instead of $ a $.

Thus, the Painleve equations form the following hierarchy:

\vspace{1pt} \ \ \ \ \ \ \ \ \ \ \ \ \ \ \ \ {P4}\ \ \ \ \ \

\ \ \ \ \ \ \ \ \ \ \ \ \ $\nearrow \quad \searrow $

P6 $\longrightarrow $ P5 \ \ \ \ \ \ \ \ \ \ \ \  {P2}
$\longrightarrow$
P1

\ \ \ \ \ \ \ \ \ \ \ \ $\searrow $\ \ \ \ \ \ \ $\nearrow $

\ \ \ \ \ \ \ \ \ \ \ \ \ \ \ \ P3 \

The  hierarchies

\vspace{1pt} \ \ \ \ \ \ \ \ \ \ \ \ \ \ \ \ {H4} \ \ \ \ \ \

\ \ \ \ \ \ \ \ \ \ \ \ \ $\nearrow \quad \searrow $

H6 $\longrightarrow $ H5 \ \ \ \ \ \ \ \ \ \ \ \ H2 $\longrightarrow$
H1

\ \ \ \ \ \ \ \ \ \ \ \ $\searrow $\ \ \ \ \ \ \ $\nearrow $

\ \ \ \ \ \ \ \ \ \ \ \ \ \ \ \ H3 \

\vspace{1pt} \ \ \ \ \ \ \ \ \ \ \ \ \ \ \ \ L4 \ \ \ \ \ \

\ \ \ \ \ \ \ \ \ \ \ \ \ $\nearrow \quad \searrow $

L6 $\longrightarrow $ L5 \ \ \ \ \ \ \ \ \ \ \ \ L2 $\longrightarrow$
L1

\ \ \ \ \ \ \ \ \ \ \ \ $\searrow $\ \ \ \ \ \ \ $\nearrow $

\ \ \ \ \ \ \ \ \ \ \ \ \ \ \ \ L3 \

form also \cite{OKO} Hamiltonians $ H1-H6 $ of the Painleve equations and
the linear ordinary differential equations of the second order $ L1-L6 $ on
an auxiliary parameter $ {\lambda} $ found in \cite{FUX}, \cite{GAR}, which
permit to investigate asymptotic behavior of the equations $ PJ $ with the
help of monodromy preserving method.

In the present work we call attention to natural and usefulness relation of
the Painleve equations  with a hierarchy

\vspace{1pt} \ \ \ \ \ \ \ \ \ \ \ \ \ \ \ \ P$4_{0}$\ \ \ \ \ \

\ \ \ \ \ \ \ \ \ \ \ \ \ $\nearrow \quad \searrow$

P$6_{0} \longrightarrow $ P$5_{0}$ \ \ \ \ \ \ \ \ \ \ \ \ {P}$2_{0}$ $\longrightarrow$
P$1_{0}$

\ \ \ \ \ \ \ \ \ \ \ \ $\searrow $\ \ \ \ \ \ \ $\nearrow $

\ \ \ \ \ \ \ \ \ \ \ \ \ \ \ \ P$3_{0}$ \

The following algebraic equations were obtained from the equations $ P1-P6
$ by a neglect by terms containing derivative
\begin{eqnarray*}
P1_{0}:&&6w^2+x=0, \\ P2_{0}:&&2w^3+xw+a=0,\\
P3_{0}:&&(aw^2+b)/x+cw^3+d/w=0,\\
P4_{0}:&&3w^2/2+4xw^2+2(x^2-a)u^2+b/w=0,\\
P5_{0}:&&(w-1)^2(aw+b/w)/x^2+cw/x+3w(w+1)/(w-1)=0,\\
P6_{0}:&&w(w-1)(w-x)[a+bx/w^2+c(x-1)/(w-1)^2+ \\
&&dx(x-1)/(w-x)^2]x^2(x-1)^2=0.
\end{eqnarray*}

2. It is well known, that it is possible to consider the Painleve
transcendets as nonlinear analogs of special functions, allowing integrated
representations of the Fourier type . Arising in applications at exposition
of various fast transients, transcendents of the Painleve allow, due to a
possibility of application to they of the monodromy preserving method,
similarly to the special functions, effectively to solve problems of
exposition of asymptotics at various values of argument $x$ \cite{JIM},
\cite {NOV}.

In view of this analogy the problem of exposition of uniform asymptotics
for the Painleve transcendents is actual (that is, about them asymptotics
also at $ x^{2} $ + $ a^{2} $ + $ b^{2} $ + $ c^{2} $ + $ d^2 $ $ \to $ $
{\infty} $), effectively solved in case of special functions, supposing
integrated representation. (The first step in this direction was actually
made recently in works by A.A. Kapaev \cite{KI}, \cite{KII}, devoted to an
asymptotics of a common solution of the equation $ P2 $ at
$\Re(a)\to\infty$ for any value of $x$. In spite of their importance, we
shall underline at once, that in these works the speech nevertheless goes
about exposition pointwise, instead of uniform on $ x $ asymptotics.)

First of all just in view of this problem alongside with a hierarchy of
equations the Painleve, it is reasonable to separate reviewing of the
entered above appropriate hierarchy of the algebraic equations $ P1_{0}
-P6_{0} $. Their solutions define the most simplest asymptotics of the Painleve
transcendents. (We shall mark also, that until now in most of all
mathematical physics applications arose the Painleve transcendents only
extremely in relation just with such asymptotics.)

All terms of this hierarchy of the algebraic equations can be
consecutively obtained from each other with the help of the same
replacements, that were described in section 1 for terms of a
hierarchy ordinary differential equations $ P1-P6 $.

Thus a quadratic equation $ P1_{0} $ is the canonical equation of a
fold type, defining in a terminology of the theory of catastrophes
\cite{GIL}, arises from each of the stayed algebraic equations of a
hierarchy in an exactitude at those values $ x $ and parameters $a$, $
b$, $ c$, $d$ (forming in aggregate gang {\it controlling} of
parameters of catastrophes circumscribed by the equations $ PJ_{0} $),
at which happens confluence of two roots of the appropriate equation $
PJ_{0} $ (in a situation of "the general provisions" multiple roots
are absent). The solution $ P1_{0} $ describes processes these
confluence, being not limited immediately by moment of degeneration
(to which in the equation $ P1_{0} $ only a value $ x = 0 $), but
includes exposition of the appropriate reorganizations, when "... the
parameter, varying, passes through a degenerate value " \cite{ARN}.

Similarly, the cubic equation $ P2_{0} $, defining canonical equation of
the cusp catastrophe, arises from each of higher on to ratio to it of the
algebraic equations $ P3_{0} $ - $ P6_ {0} $ hierarchies when three various
roots stick together.

The equations $ P3_{0} $ and $ P4_{0} $  turn out from the
equations $ P5_{0} $, $ P6_{0} $ when four various radicals stick
together.

Extreme the left arrow of a hierarchy $ P1_{0} -P6_{0} $ corresponds the
confluence of five solutions of an equation $ P6_{0} $, and from it in an
outcome of a passage to the limit there is an equation $ P5_{0} $.

The derivative of the solutions for the given hierarchy of the
algebraic equations will be tend to infinity in points, in which they
have the multiple roots. Therefore it becomes untrue a neglect terms
with derivative in small neighborhoods of such points at exposition
the Painleve transcendents assigned in a principal order to the
appropriate solutions $ PJ_{0} $. Just therefore in small
neighborhoods of such points there is essential a hierarchy,
circumscribed in section 1, of degenerations the Painleve
transcendents.

It is naturally  to consider this hierarchy as a hierarchy of
nonlinear special functions of catastrophes (SFC) of a hierarchy,
circumscribed by the equations $ PJ_{0} $.

Note 1. Similarly to the linear analogs (solutions of such equations, as,
for example, equation Bessel, Weber - Hermit, Whittaker and hypergeometric
equation of the Gauss \cite{ITS}) all solutions of the Painleve equations
simultaneously are solutions of difference equations by a variable $ a $, $
b $, $ c $, $ d $ \cite{LUK} on all from their included in right hand side.
Per last years the subjects devoted to these equations in a combination to
their continual limits, being lower terms of a hierarchy the Painleve
transcendents has become popular. The popularity of this subjects was
stimulated in the large degree by series of works of a beginning of the
90-th years under the quantum theory of a gravitation \cite{DOU}
-\cite{PER}. We mark here, that similarly to degenerations of a hierarchy
of the Painleve equations, all these continual limits are carried out also
during such small modification controlling parameters of catastrophes,
circumscribed by the algebraic equations $ PJ_{0} $, at which their
solutions pass through singularities.

3. It is traditional \cite{KIT}, the Painleve equations  are
considered as nonlinear analogs known of the linear theory (SFC),
whose integrated representations are similar to a behaviour in
neighborhoods of singular points $ \lambda $ of solutions of simple
equations of the monodromy preserving method.

The treatment, offered in the previous section, the Painleve
transcendents as is essential nonlinear special functions is based on
essence other principle ascending to variant \cite{KUD} of the
approach to nonlinear SFC, which initially being as solutions of the
partial differential equations, (such, as the Korteveg-de Vries,
Burgers and Nonlinear Schrodinger equations). The traditional
treatment of the Painleve transcendents as analogs of the Fourier
integrals was used in offered earlier \cite{ZAM}, \cite{HAB} variant
of the approach to such SFC.

It is curious, that in all considered until now examples this variant and
variant offered in \cite{KUD} "commuted" in the sense that proceeding from
different principles, invariable associated with the same algebraic
equation, defining the appropriate catastrophe.

Thus, for example, that fact, that associated with the canonical
equation of catastrophe of cusp
\bb
x-tv+v^3=0
\label{CUS}
\ee
\par
known special Gurevich-Pitaevskii solution (GP) \cite{GUR},
\cite{PIT} of the Korteveg-de Vries equation
\bb
v_{t}+vv_{x}+v_{xxx}=0
\label{KDV}
\ee
\par
is simultaneously solution of the ordinary differential equation
$$ v_{xxxx} + 5vv_{xx} /3 + 5v_{x}^2/6 + 5(x-tv + v^3) /18 = 0, $$

initially appeared in \cite{BIS} proceeding from that
circumstance, that the GP solution \cite{GUR} is analog of
special function for the cusp catastrophe:
$$
J=\int_{R}\lambda\exp(-2i(x\lambda+4t\lambda^3+3456\lambda^7/35))d\lambda
$$
( J satisfies of a linear part of (\ref{KDV}), and its
asymptotics at $ x^2 $ + $ t^2 $ $ \to $ $ \infty $ following to
a method of a stationary phase \cite{F}, is defined in terms of
solutions of (\ref{CUS}.)

In \cite{KUD} was specified, that the validity for a special GP solution of
the given equation could be deduced from that fact, that at a rejection in
it the derivative there is in an exactitude a solution of the cusp
equation
(\ref{CUS}), which defines \cite{GUR} an asymptotics on infinity of a
solution GP outside of area of its fast oscillations.

Let's mark, that similar "commutation" of two treatments of the Painleve
equations as nonlinear SFC has a place of the equations $ P1 $ and $ P2 $.
Really, according to traditional treatment the equation $ P1 $ is nonlinear
analog of the Fourier integral
$$
\int_{R} \exp(ix\lambda + i\lambda^5) d\lambda,
$$
and equation $ P2 $ is the analog of an integral
$$
\int_{R} \lambda^{ia} \exp(-ix\lambda + i\lambda^4) d\lambda.
$$
As on a method of a stationary phase an asymptotics of first of
them at $x\to  \infty $ can be expressed in terms of a solution
for the equation $P1_{0}$, and the asymptotics of second integral
at $ x^2+ a^2\to\infty $ in terms of solutions $ P2_{0} $, at
both treatments the first Painleve is considered in quality SFC
such as canonical fold catastrophe, and second such as cusp
catastrophe.

However for the remaining Painleve equations  the fact of similar
"commutation" is not traced, probably, truth, only at present. But, in
any case, in an association from a context of a problem, in which
there can be that or other Painleve equation, it is useful to mean
both treatments.

4. That fact, that in process of the Óonfluence of the roots of a
hierarchy of the algebraic equations $ PJ_{0} $ at exposition of
asymptotics of the equations $ PJ $ is necessary to use the solutions
of the Painleve equations $ PK $ with numbers $ K $ $<$ $ J $ does not
solve completely problem on exposition of a behaviour of these
solutions of the equations $ PJ $ in small neighborhoods of values
controlling parameters of catastrophes $ PJ_{0} $, at which happens
the Óonfluence of the various roots of this equation. The additional
analysis is necessary here. It is clear even on an example of an
outcome of degeneration of the second Painleve equation to the first
Painleve equation:

While the solutions of the second Painleve equation can not have poles
higher than first order, all real solutions of the equation $ P1 $ at
real $ x $ have an infinite set of second order poles.

The similar position has a place and for degenerations others the Painleve
equations. (It is also for degenerations of difference equations, with
which simultaneously satisfy appropriate the Painleve equations.)

Besides the analysis about influence of derivative asymptotics is necessary
and after completion of reorganization of solutions of the equations
$PJ_{0} $ during a modification controlling parameters $ x $, $ a $, $ b $,
$ c $, $ d $, during which these solutions of the algebraic equations pass
through singularities.

In the stayed sections of the paper, we shall not concern a problem on
an improvement a behaviour of solutions of the equations $ PJ $ in
neighborhoods of poles of solutions of the lowest terms of a hierarchy
of the Painleve equations, by which they are reduced during passages
to the limit, circumscribed in section 1. Up to the present moment it
remains not investigated.

Let's mark only that circumstance, that the special solutions of first
Painleve equations having at $x\to\infty $ asymptotic expansions as
ascending power series should arise similarly at exposition of
asymptotics of solutions for a wide class of the nonlinear equations
(as ordinary, and in partial derivatives) with small parameters at
derivative.

In particular, these solutions arise at exposition formal asymptotics of a
solution for singular - perturbed equation
\bb
i{\epsilon}\Psi_{t}+(|\Psi|^2-t)\Psi=1
\label{KIS}
\ee
\par
in a small neighborhood of a point $ t = T = 3/2^{1/3} $, which at $
t> T $ has asymptotic expansion as a series
$$
\Psi_{0} (t) + \epsilon\Psi_{1} (t) + \epsilon^2\Psi_{2} (t) +...,
$$
in which the principal term $ \Psi_{0} $ = $ u (t) $ is one of solutions of
a cubic equation
$$
u^3-tu=1,
$$
having a singularity in a critical point $ t = T $. The problem that
happens, for example, with an asymptotics of a solution ( \ref{KIS}) at
passage through a critical point, which first of two authors of the given
paper was set of problem, and it have reduced at the end to this preprint.

4. The rest of the given work is devoted, in basic, research of connection
between nonoscillatory and fast oscillating by asymptotics on a small
parameter $ \ve $ to the real solution of the equation:
\bb
\ve^2 \pt_t^2 u+2u^3-tu=1,
\label{p2}
\ee
by obvious replacement reducible to only imaginary (at real $ x $)
solutions of the equation $ P2 $. Asymptotic solutions of this equation as
a series
\bb
u = u_{0} + \epsilon^2 u_{1} + \epsilon^4 u_{2} +...,
\label{ROW}
\ee
where the principal term $ u_{0} $ is the solution of a cubic equation
\bb
2u^3-tu=1,
\label{p20}
\ee

This asymptotics
becomes unsuitable at passage through points of a singularity of these
solutions (after their passage $ u_{0} $ ceases to be clean by real
function).

It is necessary to take into consideration the influence of the derivative
for exposition of a principal term of an asymptotics. It is naturally for
the considered problem to expect, that there will be fast oscillating
asymptotic solutions (WKB-solutions, averaged on Kuzmak-Witham of a
solution) of the equation ( \ref{p2}), which principal term was actually
constructed by Kuzmak in him widely to known work \cite{K} (construction of
full asymptotic expansion and its justification is present in \cite{FMI}).

The results of the numerical account, reduced in section 5, confirm this
supposition for one of three solutions of a cubic equation (\ref{p20}).
\par
We will be interested by a problem on connection of the asymptotic solution
of the equation (\ref {p2}), which main term is in the root of a cubic
equation, and the oscillating asymptotic solutions for the equation
Painleve-2 is constructed by Kuzmak \cite {K}, which, as we shall assume,
are served for explanation of the results for this numerical account.

Our problem is from a two-parameter set of fast oscillating asymptotic
solutions to choose a solution, in which passes an asymptotic solution
(\ref{ROW}) after passage t through a critical value $ t = t_* $.
\par
Let's underline, that we consider this problem  in the first turn as a
model example for study of the connection of various types of
asymptotic solutions of a wide class of the singular-perturbed
nonlinear differential equations (and not only ordinary). The choice
of this equation as a model is explained by its simple form, and also
analytically any solution at real $ t $\cite{LUK}. In our further
analysis, as against works \cite{KI}, \cite{KII}, property of an
integrability of the equation $ P2 $ by the monodromy preserving
method is not used, that allows to hope on applicability of similar
reasonings and at the analysis of unintegrable problems.

Note 2. Besides the results \cite{KI},\cite {KII} give uniform asymptotics
in small ranges of an explanatory variable $ x-x_0\ll\a^{-1/3-\d} $, for
anyone as much as small $ \d > 0 $ or, accordingly $t-t_0\ll\ve^{-1-\d}$
(here $ x_0 = const $). In our work the uniform asymptotics on $ \epsilon $
are constructed in the large area (order $O(1)$) of the explanatory
variable $ t $ in natural terms for a considered problem of stretchings and
replacements. It is clear, that only use of similar stretchings and
replacements can allow to receive the satisfactory answers in unintegrable
problems.

But as one more argument for the benefit of our choice, certainly, the
connection with explained above in sections 1-3 has served also. The
equation (\ref{p2}) by the stretchings
$X=\ve^{2/3}t,\,W=\ve^{1/3}u,\,\a=\ve^{-1}$ is reduced to the equation
\bb
y_{xx}-xy+2y^3=\k.
\label{PIM}
\ee
\par
It is one of the forms of the second Painleve equation(replacement $Y=iw$,
$k=ia$, reduce it to canonical form $P2$). Outside of direct $ K = 0 $ the
problem on an asymptotics at $x^2+k^2\to\infty$ of solutions of this
equation is reduced to study an asymptotics on $\epsilon$ uniform on $ t $
of the solution (\ref{p2}).

\par
5. The asymptotic solution (\ref {ROW}) is defined completely by a choice
of the roots of the cubic equation for the principal term of the
asymptotics. There is a point $t_*$, more to the right of which the cubic
equation has three material roots $u_{00}(t)>u_{01}(t)>u_{02}(t)$. In a
point $ t= t_*$ the roots $u_{01}(t)$ and $u_{02}(t)$ stick together. The
value $T_*$ and magnitude $u_*=u_{01}(t_*)=u_{02}(t_*)$ are defined from
the equations:
$$
2u_*^3-t_*u_*=t,\quad 6u_*^2-t_*=0.
$$
Thus asymptotic solution (\ref{ROW}) with a principal term
$u(t)_{0}=u_{0i}(t),\! i=1,2$ is suitable at $t>t_*$. In the
neighborhood of the point $t_*$ such expansion (\ref{ROW}) loses the
asymptotic character because of growth of the correction terms. If as
a principal term to take $u_{00}(t)$, then the asymptotic solution
(\ref{ROW}) will be suitable at any value $t$. (This reasoning is
confirmed in the following paragraph by results of numerical
experiments).

\par
The choice of a considered below principal term of a solution (\ref{ROW})
is connected also to the problem on its stability in relation to small
perturbations. The solutions with principal terms $u_0(t)$ or $u_2(t)$ are
stable. It follows from the analysis of the equation (\ref{p2}), linearized
on the principal term of the formal asymptotics:
$$
\ve^2\pt_t^2 v+(6\stackrel{0}{u}\!^2(t)-t)v=0.
$$
Therefore for the analysis of connection between an asymptotic solution
(\ref{ROW}) of a dispersionless limit and oscillating asymptotic solutions
at $t>t_*$ we shall choose an asymptotic solution in the form (\ref{ROW})
where the principal term is $\stackrel{0}{u}(t)=u_2(t)$. At first, - it is
stable, secondly, - it is not suitable at $t<t_*$. More to the left of the
point $t_*$ we shall construct fast oscillating asymptotic solution, which
is in turn unsuitable at $t>t_*$ and its principal term continuously passes
in a leading term asymptotic solution (\ref{ROW}).

Let's reduce outcomes of a numerical solution by a Runge-Kutta method of a
Cauchy problem for (\ref {p2}) with initial data -, appropriate to the
three roots of the equation (\ref{p20}).

\includegraphics[width=100mm,angle=0]{p-21.ps}\\
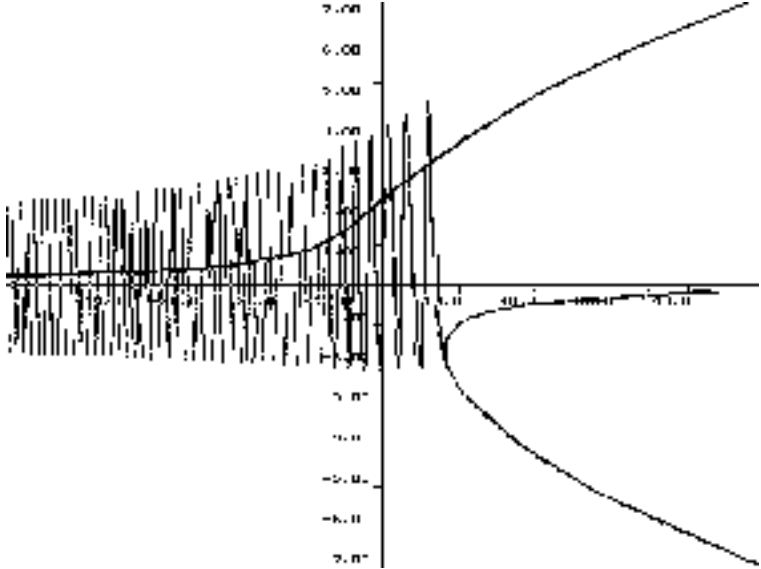
\centerline{Fig.1}
\vspace*{3mm}

\par
>From the figure 1 it is obvious, that after passage through a point $
t_* $ on the right to the left the solution becomes fast oscillating.
The process of origin of these oscillations represented on figure 1 is
pertinent, using analogies from the theory of bifurcations \cite{ARN},
to name to a rigid regime of establishment of oscillations. As it is
visible from this figure at $ a\to-\infty $, these the amplitude of
these oscillations aspires to zero. In a combination to results of
numerical experiment reduced on Figure 1 these circumstances allows to
put forward a hypothesis that in a neighborhood direct $ k = 0 $ in an
asymptotics of a solution of the equation (\ref{PIM})  $ Y $ at $ x^2
+ k^2\to\infty $ qualitative reorganization (which, again using
arising analogies to a standard terminology of the theory of
bifurcations also happens, it is natural to name as a soft regime of
origin of oscillations).

Unfortunately, even in case of a validity of this hypothesis, solution
of the important problem of deriving uniform asymptotics $ y $ at
$T\to-\infty$ on the present moment is not obtained. We shall be
forced to be limited to exposition of this reorganization at a level
of presentation in spirit \cite{KI},\cite{KII} asymptotics $ y $ at $
t \to-\infty $ for anyone fixed $ k $:

At $ k < 0 $ it carries an algebraic character and easily it
turns out by approaching replacements from a series (\ref{ROW});

At $ k = 0 $ the principal term of an asymptotics $ y $ is obtained in
recent work \cite{BEL} (exposition after that full asymptotic
expansion turns out with the help of standard reasonings)
$$
Y = d (-t)^{-1/4} {\sin (2/3 (-t)^{3/2} -3d^2ln (-t) /4 + \beta},
$$
with some constants $d$,$\beta$, which are calculated in \cite{BEL};

And at $ k> 0 $ the answer is described with the help by an fast
oscillating asymptotic solution from units 7 of the present work. (From
results of unit 7 follows, that the considered there asymptotic solution
aspires to zero at $ t \to\infty$.

\par
6. The special character of direct $ k = 0 $ for considered by us SFC of
cusp, which in this case is described with the help of the solutions of the
cubic equation
\bb
2y^3-xy-k = 0,
\label{KUB}
\ee
\par
carries a not casual character. It forms so-called {\it Maxwell
stratum} of cusp catastrophes. (From a point of view of the
catastrophes theory the Maxwell stratum what is selected because on it
the various radicals (\ref{KUB}) define the same value of a primitive
$ y^4/2-xy^2/2-ky $ it was specified of the left part of this cubic
equation.) It was specified already in \cite {BIS}, the special
character of  the Maxwell stratum one can find in the asymptotic
expansions of a lot various linear and nonlinear SFC associated with
catastrophe of cusp.

We would like to pay attention here for the following:

In one of papers V.R.Kudashev and second of the authors of the present
paper preparing in the present moment for the publication, it is marked,
that for a some nonlinear SFC (equations, being simultaneously solutions of
Burgers and Nonlinear Schrodinger equations) the Maxwell stratum is
selected not only at a level of asymptotic expansions:

for these SFC on it the "increased" integrability has a place. The
ordinary differential equations, which (alongside with initial
integrable partial equations) the data nonlinear special functions
satisfy, because of their parity, or the oddness, suppose on the
Maxwell stratum lowering of the order.

There is, a property of the "increased" integrability on the Maxwell
stratum  is not limited at all to cases SFC, which are initially solutions
of integrable equations and follows from their parity or oddness not
necessarily. Really, ordinary differential equation (\ref {KIS}) is reduced
to the equation
$$
iQ _ {z} + (| Q |^2-z) Q = a,
$$
which at $ a = 0 $ is integrable in quadratures.

One more similar example give SFC of the cusp catastrophe being a solution
of the Abel equation:
$$
Q_{z}=Q^{3}-zQ+a.
$$

7. For clearing up of a qualitative character of a behaviour of a
(\ref{p2}) solution, interesting for us, after passage through $ t _ *
$ the following reception is applicable. By "freezing" a value of
factor $ t $ in the equation (\ref{p2}) in some point $ t = T $, we
shall consider equation
$$
\ve^2\pt_t^2 V+2V^3-TV=1.
$$
Integrating it one, we shall receive a relation:
$$
-\ve^2(\pt_tV)^2=V^4-TV^2-2V-E.
$$
In an association from a value of the constant $ T $ the potential has
one of the curve:

\includegraphics[width=100mm,angle=0]{p-22.ps}\\
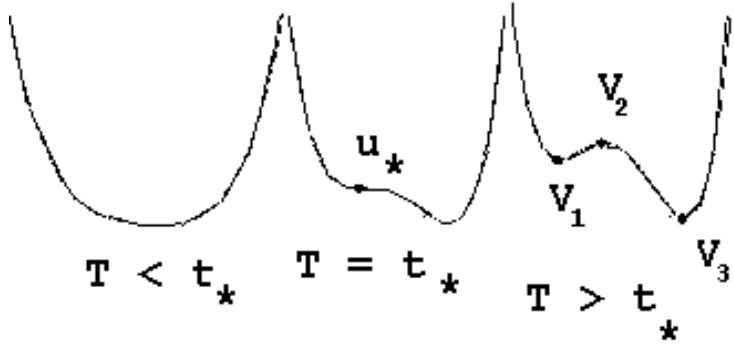
\centerline{Fig.2}

At $T>t_*$ the points $V_0$ and $V_2$ are the point of a stable
equilibrium, and $ V_1 $ is the point of a labile equilibrium. It is
easy to see, that at $t=T$ the value of a principal term of an
asymptotic solution (\ref{ROW}) (the function $ u_2 (t) $) coincides
with $V_2$. At $T=t_*$ the points $ V_1 $ and $ V_2 $ stick together
and $ u_* $ is the point of the labile equilibrium. In particular, in
this point the value of a parameter $ E = E_* $ is easy for
calculating for an equilibrium condition:
$$
E_*=u_*^4-t_*u_*^2-2u_*.
$$
\par
The sequence the picture at $ T> t_*, \quad T = t_*, \quad T < t_* $ gives
the qualitative answer to the problem: that happens to a true solution of
the equation (\ref{p2}) more to the left of a point $ t_* $. The solution
carries an oscillatory character. The value of a parameter $ E $ - analog
an energy - in a point $ t_* $ is equal to $ E_* $.

\par
8. Let's pass to a construction of an asymptotics of this oscillating
solution. Following \cite{F} we shall search for it as:
\bb
u(t,\ve)=\stackrel{0}{U}(t_1,t,\ve)+\ve\stackrel{1}{U}(t_1,t,\ve)+\dots.
\label{exp2}
\ee
As argument $ t_1 $ we shall use the expression $ S (t) /\ve $, where
$S(t)$ is unknown function. The equations for the definition of an
association from a parameter $ t_1 $ of first two terms of an
asymptotics look like:
$$
(S')^2\pt_{t_1}^2\stackrel{0}{U}+2\stackrel{0}{U}\!^3-\stackrel{0}{U}t=1,
$$

$$
(S')^2\pt_{t_1}^2\stackrel{1}{U}+(6\stackrel{0}{U}\!^2-t)\stackrel{1}{U}=
-2S'\pt^2_{tt_1}\stackrel{0}{U}-S''\pt_{t_1}\stackrel{0}{U}.
$$
Integrate once on $ t_1 $ the equation for $ \stackrel{0}{U} $ in an
outcome we get:
\bb
(S')^2(\pt_{t_1}\stackrel{0}{U})^2=-\stackrel{0}{U}\!^4+
t\stackrel{0}{U}\!^2+2\stackrel{0}{U}+E(t),
\label{eq-U}
\ee
where $ E (t) $ is "the constant of integration". In \cite{K} it is shown,
that the condition of periodicity on a parameter $ t_1 $ for the function
$\stackrel{1}{U} $ reduces in the equation for function $ S (t) $:
$$
S'\int_0^T\big[\pt_{t_1}\stackrel{0}{U}(t_1,t)\big]^2dt_1=c_0.
$$
Here $ T $ is period of the oscillations, $ c_0 $ - constant. By
taking in account the explicit expression for the derivative on $ t_1
$ this formula can be copied in a little bit other form:
\bb
2\int_{\b(t)}^{\a(t)}\sqrt{-x^4+tx^2+2x+E(t)}dx=c_0,
\label{eq-S}
\ee
where $ \a (t) $ and $ \b (t) $ are the solutions of the equation $ -x^4 +
tx^2 + 2x + E (t) = 0 $. For the present the function $ E (t) $ is not
defined here. Its connection with a phase of the fast oscillations $ S (t)
$ is given by the formula \cite{F}:
\bb
T=\sqrt{2}S'\int_{\b(t)}^{\a(t)}{dx\over\sqrt{-x^4+tx^2+2x+E(t)}}.
\label{eq-E}
\ee
{ \bf Note 3.} A.N. Belogrudov has specified to the authors, that the
integral in the left part (\ref{eq-S}) is generalized hypergeometric
function satisfying to a set of equations in partial derivatives on
parameters $ \a $ and $ \b $ \cite{GR}.
\par
The equations (\ref{eq-U}) - (\ref{eq-E}) define to within some constant $
c_0 $ the principal term of the asymptotics (\ref{exp2}). Let's remark,
that we build the asymptotic solution of the equation (\ref{p2}) at $ t <
t_* $. Thus the polynomial of the fourth degree on $ \stackrel{0}{U} $ in a
right hand side of the equation (\ref{eq-U}) can have no more two various
real roots $ \a (t) $ and $ \b (t) $. Hence, this polynomial can be look
as:
$$
F(x,t)=(\a(t)-x)(x-\b(t))\Big((x-m(t))^2+n^2(t)\Big).
$$
\par
In the point $ t = t_* $ the curve on the figure 2 has a point of
inflection. The degeneration of an elliptic integral at $ t = t_* $
corresponds to a case $ M (t_*) = \b(t_*)=u_*$ and $n(t_*)=0$, when one of
the roots of a polynomial corresponds to the value of the polynomial in a
point of inflection. For this case it is easy to calculate the constant in
the right hand side of the equation (\ref{eq-S}): $ C_0 = \pi $ and value
of the parameter $ E (t_*) = E_* $.

\par
Express gratitude L.A. Kalyakin, S.G. Glebov and A.N. Belogrudov for
discussions  and also V.E.Adler for help in realization of the numerical
calculations.
\par
The work was maintained RFBR 97-01-00459, 96-01-00382 and Fund of support
of scientific schools 96-15-96241.

\end{document}